\documentclass[prl,twocolumn,showpacs,preprintnumbers,amsmath,amssymb,superscriptaddress]{revtex4}

\usepackage{bm}
\usepackage{graphicx}
\usepackage{amsbsy}
\usepackage{amsmath}
\usepackage{amsfonts}
\usepackage{amsthm}

\begin{document}

\theoremstyle{plain}
\newtheorem{theorem}{Theorem}
\newtheorem{lemma}[theorem]{Lemma}
\newtheorem{corollary}[theorem]{Corollary}
\newtheorem{conjecture}[theorem]{Conjecture}
\newtheorem{proposition}[theorem]{Proposition}

\theoremstyle{definition}
\newtheorem{definition}{Definition}

\theoremstyle{remark}
\newtheorem*{remark}{Remark}
\newtheorem{example}{Example}

\title{Reexamination of Entanglement of Superpositions}   
\author{Gilad Gour}\email{gour@math.ucalgary.ca}
\affiliation{Institute for Quantum Information Science and 
Department of Mathematics and Statistics,
University of Calgary, 2500 University Drive NW,
Calgary, Alberta, Canada T2N 1N4} 

\date{\today}

\begin{abstract} 
We find tight lower and upper bounds on the entanglement of a superposition
of two bipartite states in terms of the entanglement of the two states constituting
the superposition. Our upper bound is dramatically tighter than the one presented
in Phys. Rev. Lett \textbf{97}, 100502 (2006) and our lower bound can be used to 
provide lower bounds on different measures of entanglement such as the 
entanglement of formation and the entanglement of subspaces . 
We also find that in the case in which the two states are one-sided orthogonal, the entanglement of the superposition state can be expressed explicitly in terms of the entanglement of the two states in the superposition.
\end{abstract}  


\maketitle

In a recent paper by Linden, Popescu and Smolin (LPS)~\cite{Lin06}, 
the authors raised the following question: Given a \emph{bipartite} state 
$|\Gamma\rangle$, and given a certain 
decomposition of it as a superposition of two bipartite states
$$
|\Gamma\rangle=\alpha|\Psi\rangle+\beta|\Phi\rangle\;;\;\;\;|\alpha|^2+|\beta|^2=1
$$ 
what is the relation between the entanglement of $|\Gamma\rangle$ and the entanglement
of $|\Psi\rangle$ and $|\Phi\rangle$? It is somewhat surprising that very little is known
about this basic question given how important entanglement is to quantum mechanics and 
how in bipartite settings superposition is almost a synonymous term to entanglement.
Perhaps one of the reasons for that is that the entanglement of 
$|\Gamma\rangle$, 
depends also on the coherence between the two terms in the decomposition, 
and therefore in general it does not depend only on the entanglement of $|\Psi\rangle$
and $|\Phi\rangle$. This can be seen most clearly in the Bell state example with
$|\Psi\rangle=|00\rangle$, $|\Phi\rangle=|11\rangle$, and $\alpha=\beta=1/\sqrt{2}$.
Nevertheless, in~\cite{Lin06} the authors have found an upper bound 
(dubbed here the LPS bound) on 
the entropy of entanglement of $|\Gamma\rangle$ given in terms of the entanglement 
of $|\Psi\rangle$ and $|\Phi\rangle$. Subsequently, several authors generalised this result 
to include different measures of entanglement~\cite{Cha07,Yon07,Nis07}, 
entanglement of superpositions of multipartite states~\cite{Cav07,Son07} and 
entanglement superpositions of more than two states~\cite{Yan07}. 

In this Letter we show that the LPS upper bound is not tight and 
can be improved dramatically if one
includes two factors. The first one is based on a generalization of biorthogonal
states to include one-sided orthogonal bipartite states. This factor leads to a 
slight improvement of the LPS bound. The second more important factor that
leads to a dramatic improvement is based on the relation between different 
convex decompositions of a density matrix. 
We find that unless $E(\Psi)=E(\Phi)$ and $|\alpha|=|\beta|$ our bound is strictly tighter
and in general, in the limit of large dimensions can be arbitrarily tighter.  Our method also
enables us to find a tight lower bound that depends only on $E(\Psi)$, $E(\Phi)$, 
$|\alpha|$ and $|\beta|$. 

We start with a definition of one-sided orthogonal bipartite states and a simple
improvement of the LPS bound. 
\begin{definition} \emph{One sided orthogonal bipartite states:} Two bipartite states
$|\Psi\rangle_{AB}$ and $|\Phi\rangle_{AB}$ are one sided orthogonal if 
\begin{equation}
\text{Tr}_{B}\left[\text{Tr}_A\left(|\Psi\rangle\langle\Psi|\right)
\text{Tr}_A\left(|\Phi\rangle\langle\Phi|\right)\right]=0
\label{sided1}
\end{equation}
or
\begin{equation}
\text{Tr}_{A}\left[\text{Tr}_B\left(|\Psi\rangle\langle\Psi|\right)
\text{Tr}_B\left(|\Phi\rangle\langle\Phi|\right)\right]=0\;.
\label{sided2}
\end{equation}
\end{definition}
Note that one sided orthogonal states are orthogonal but not necessarily 
biorthogonal (i.e. for one sided orthogonal states in general only one of
the two equations above is satisfied). In the following, with out loss of generality,
we assume that one-sided orthogonal states satisfy Eq.~(\ref{sided1}) but not necessarily
Eq.~(\ref{sided2}).

\begin{lemma}
Up to local unitary transformations, one-sided orthogonal states can be written
as:
\begin{equation}
|\Psi\rangle=\sum_{i=1}^{d_1}\sqrt{p_i}|u_i\rangle_A|i\rangle_B\;\;\text{and}\;\;
|\Phi\rangle=\sum_{i=1}^{d_2}\sqrt{q_i}|v_i\rangle_A|i+d_1\rangle_B\;,
\label{lemma}
\end{equation}
where $\{p_i\}$ and $\{q_i\}$ are two sets of positive numbers that sums to one,
and $\{|u_i\rangle_A\}$ and $\{|v_i\rangle_A\}$ are two sets of orthonormal states.
\end{lemma}
Note that if ${}_A\langle v_{i'}|u_i\rangle_A=0$ for all $i=1,2,...,d_1$ and $i'=1,2,...,d_2$
then the states are biorthogonal.
\begin{proof}
Due to the Shmidt decomposition we have
$$
|\Psi\rangle=\sum_{i=1}^{d_1}\sqrt{p_i}|u_i\rangle_A|u_i\rangle_B\;\;\text{and}\;\;
|\Phi\rangle=\sum_{i=1}^{d_2}\sqrt{q_i}|v_i\rangle_A|v_i\rangle_B\;,
$$
where $\{|u_i\rangle\}$ and $\{|v_i\rangle\}$ are sets of orthonormal states,
and $\{p_i\}$ and $\{q_i\}$ are two sets of positive numbers that sums to one.
Since we assume that the states satisfy Eq.~(\ref{sided1}), we get that
${}_B\langle v_{i'}|u_i\rangle_B=0$ for all $i=1,2,...,d_1$ and $i'=1,2,...,d_2$. Thus,
we can define the set $\{|i\rangle_B\}_{i=1}^{d_1+d_2}$, where 
$|i\rangle_B=|u_i\rangle_B$ for $1\leq i\leq d_1$ and $|i\rangle_B=|v_{i-d_1}\rangle_B$ for 
$d_1+1\leq i\leq d_1+d_2$. With these notations we obtain Eq.~(\ref{lemma}).
\end{proof}

\begin{theorem}
Given $|\Psi\rangle$ and $|\Phi\rangle$ one-sided orthogonal, 
and $|\alpha|^2+|\beta|^2=1$, the entanglement of the superposition
$|\Gamma\rangle=\alpha|\Psi\rangle+\beta|\Phi\rangle$ obeys
\begin{align}
E(\Gamma) = S(\rho^A) &=|\alpha|^2E(\Psi)+|\beta|^2E(\Phi)\nonumber\\
& +S(\rho^{AB})-|S(\rho^A)-S(\rho^B)|
\label{one-sided}
\end{align}
where $\rho^{AB}=|\alpha|^2|\Psi\rangle\langle\Psi|+|\beta|^2|\Phi\rangle\langle\Phi|$
and $\rho^A$ and $\rho^B$ are obtained by tracing $\rho^{AB}$ over B and A, respectively.
\end{theorem}
Few remarks are in order. First, since $|\Phi\rangle$ and $|\Psi\rangle$ are orthogonal
we have $S(\rho^{AB})=h_2(|\alpha|^2)$ where $h_2(x)=-x\log x-(1-x)\log(1-x)$ is the 
binary entropy function. Second, for biorthogonal
states $S(\rho^A)=S(\rho^B)$ and so we obtain the formula given in~\cite{Lin06} for 
that case. Third, note that the right hand side of Eq.~(\ref{one-sided}) depends
only on quantities with no coherence between $|\Phi\rangle$ and $|\Psi\rangle$. 
Forth, from the triangle inequality of the von-Neumann entropy
(i.e. the Araki-Lieb inequality) we have $S(\rho^{AB})\geq |S(\rho^A)-S(\rho^B)|$. 

\begin{proof}
Due to Lemma~1 we have 
$\text{Tr}_B|\Gamma\rangle\langle\Gamma|=\rho^A$. Hence, $E(\Gamma)=S(\rho^A)$.
Further, from Lemma~1 it follows that the eigenvalues of
$\rho^B\equiv |\alpha|^2\text{Tr}_A|\Psi\rangle\langle\Psi|
+|\beta|^2\text{Tr}_A|\Phi\rangle\langle\Phi|$
are $\{|\alpha|^2p_i\}$ and $\{|\beta|^2q_j\}$ for $i=1,...,d_1$ and $j=1,...,d_2$. 
Thus, $S(\rho^B)=|\alpha|^2E(\Psi)+|\beta|^2E(\Phi)
+S(\rho^{AB})$. This also implies that for one-sided orthogonal states
satisfying  Eq.~(\ref{sided1}), $S(\rho^B)\geq S(\rho^A)$. This completes the proof.
\end{proof}

\begin{example}
Consider the one-sided orthogonal states 
$$
|\Psi\rangle=\frac{1}{\sqrt{2}}(|0\rangle|0\rangle+|1\rangle|1\rangle)\;\;\text{and}\;\;
|\Phi\rangle=\frac{1}{\sqrt{2}}(|0\rangle|2\rangle+|1\rangle|3\rangle)\;.
$$
Clearly, $E(\Psi)=E(\Phi)=1$. Now, it is also easy to check that $E(\Gamma)=1$
for \emph{any} coherent superposition 
$|\Gamma\rangle=\alpha|\Psi\rangle+\beta|\Phi\rangle$. Therefore, the left hand side
of Eq.~(\ref{one-sided}) is equal to 1. One can also check that $S(\rho^A)=1$ whereas
$S(\rho^B)=1+h_2(|\alpha|^2)$. Thus, the right hand side of Eq.(\ref{one-sided}) is 
also equal to 1.
\end{example}

Before we present our two main results (Theorem 3 and Theorem 4), 
we briefly review the main theorem in~\cite{Lin06} and provide a slight 
improvement of it. The authors in~\cite{Lin06} 
have used two properties of the 
von-Neumann entropy:
\begin{equation}
|\alpha|^2S(\sigma_1)+|\beta|^2S(\sigma_2)\leq 
S(|\alpha|^2\sigma_1+|\beta|^2\sigma_2)
\label{concave}
\end{equation} 
and
\begin{equation}
S(|\alpha|^2\sigma_1+|\beta|^2\sigma_2)\leq
|\alpha|^2S(\sigma_1)+|\beta|^2S(\sigma_2)+h_2(|\alpha|^2).
\label{conc}
\end{equation}
Now, consider $\rho^{AB}$ and $\rho^A$ as defined in Theorem 1, except 
that now $\Psi$ and $\Phi$ are not necessarily orthogonal.
We can write $\rho^A=|\alpha|^2\sigma_1+|\beta|^2\sigma_2$, where
$\sigma_{1}=\text{Tr}_B|\Psi\rangle\langle\Psi|$ and 
$\sigma_{2}=\text{Tr}_B|\Phi\rangle\langle\Phi|$. Using Eq.~(\ref{conc})
we get,
\begin{equation}
S\left(\rho^A\right)\leq|\alpha|^2E\left(\Psi\right)+|\beta|^2E\left(\Phi\right)
+h_2\left(|\alpha^2\right)\;.
\label{lps1}
\end{equation}
Next, the state $\rho^A$ can also be decomposed as 
$\rho^A=(n_{+}/2)\sigma_{+}+(n_{-}/2)\sigma_{-}$, where 
$\sigma_{\pm}\equiv (1/n_{\pm})\text{Tr}_B|\Gamma_{\pm}\rangle\langle\Gamma_{\pm}|$,
$|\Gamma_{\pm}\rangle=\alpha|\Psi\rangle\pm\beta|\Phi\rangle$ and 
$n_{\pm}=\| |\Gamma_{\pm}\rangle\|^2$.
Thus, from Eq.(\ref{concave}) we get
\begin{equation}
\frac{n_{+}}{2}E(\Gamma_+)+\frac{n_{-}}{2}E(\Gamma_-)\leq S\left(\rho^A\right)\;,
\label{lps2}
\end{equation}
(the notation $E(\Gamma_{\pm})$ refers to the entanglement of the normalized
states $(1/\sqrt{n_{\pm}})|\Gamma_{\pm}\rangle$).
Combining Eq.~(\ref{lps1}) with Eq.~(\ref{lps2}) and using the fact that 
$E(\Gamma_-)\geq 0$ one obtains the LPS bound:
$$
\|\Gamma\|^2 E\left(\Gamma\right)\leq 2\left( |\alpha|^2E\left(\Psi\right)+|\beta|^2E\left(\Phi\right)
+h_2\left(|\alpha^2\right)\right),
$$
(here $\Gamma\equiv\Gamma_+$).

We now present a simple improvement of the above LPS bound.
\begin{theorem}
Let $|\Psi\rangle$ and $|\Phi\rangle$ be two bipartite states, and let
$\alpha,\beta\in\mathbb{C}$ such that $|\alpha|^2+|\beta|^2=1$. Then,
\begin{align}
\|\alpha|\Psi\rangle+\beta|\Phi\rangle \|^2E(\alpha|\Psi\rangle &+\beta|\Phi\rangle) \leq 2\Big[
|\alpha|^2E(\Psi)+|\beta|^2E(\Phi)\nonumber\\
& +h_2\left(|\alpha|^2\right)-|S(\rho^A)-S(\rho^B)|
\Big]\;.\nonumber
\end{align}  
\end{theorem}
\begin{proof}
To prove it we improve the bounds given in Eqs.~(\ref{lps1},\ref{lps2}).
Eq.~(\ref{lps2}) can be slightly improved by writing
\begin{equation}
\frac{n_{+}}{2}E(\Gamma_+)+\frac{n_{-}}{2}E(\Gamma_-)\leq 
\min\{S\left(\rho^A\right),S\left(\rho^B\right)\}\;,
\label{t21}
\end{equation}
since one can repeat the same arguments that led to Eq.~(\ref{lps2}) with $\rho^B$ 
instead of $\rho^A$.
In the same way, Eq.~(\ref{lps1}) can be improved to the following one:
$$
\max\{S(\rho^A),S(\rho^B)\}\leq|\alpha|^2E\left(\Psi\right)+|\beta|^2E\left(\Phi\right)
+h_2\left(|\alpha^2\right).
$$
Thus,
\begin{align}
\min\{S(\rho^A),S(\rho^B)\} & \leq|\alpha|^2E\left(\Psi\right)
+|\beta|^2E\left(\Phi\right)\nonumber\\ 
&+h_2\left(|\alpha^2\right) -|S(\rho^A)-S(\rho^B)|.
\label{t22}
\end{align}
The combination of Eq.(\ref{t21}) and Eq.~(\ref{t22}) provides the proof for 
Theorem 2.
\end{proof}

\begin{example}
Consider the following example when Alice and Bob have Hilbert 
spaces of dimensions 3 and 4, respectively: 
\begin{align}
|\Psi\rangle & =\sqrt{\frac{1}{2}}|00\rangle+\frac{1}{2}|11\rangle
+\frac{1}{2}|22\rangle\nonumber\\
|\Psi\rangle & =\sqrt{\frac{1}{2}}|03\rangle+\frac{1}{2}|11\rangle
+\frac{1}{2}|22\rangle\nonumber\\
\alpha & =\beta=\frac{1}{\sqrt{2}}\;.\nonumber
\end{align}
The entanglement of $|\Psi\rangle$ and $|\Phi\rangle$ is $3/2$,and the entanglement
of $\alpha|\Psi\rangle+\beta|\Phi\rangle$ is $\log 3>3/2$. Since 
$\| \alpha|\Psi\rangle+\beta|\Phi\rangle\|=\sqrt{3/2}$, the LPS upper bound
is $E(\alpha|\Psi\rangle+\beta|\Phi\rangle)\leq 5\sqrt{2/3}$. 
Since $S(\rho^A)=3/2$ and $S(\rho^B)=2$, our bound is
$E(\alpha|\Psi\rangle+\beta|\Phi\rangle)\leq 4\sqrt{2/3}$ (i.e. an improvement
by almost 1 ebit). 
\end{example}

The bound in theorem 2 provides an improvement of the LPS bound. However, 
since $h_2(|\alpha|^2)\geq |S(\rho^A)-S(\rho^B)|$ our bound is smaller
by no more than 2$ebits$ from the LPS bound. We now ready to introduce 
first of our two main results which provides a new upper bound that can be arbitrarily 
smaller than the LPS bound. 

\begin{theorem}
Let $|\Psi\rangle$ and $|\Phi\rangle$ be two bipartite states, and let
$\alpha,\beta\in\mathbb{C}$ such that $|\alpha|^2+|\beta|^2=1$. Then,
\begin{equation}
\|\alpha|\Psi\rangle+\beta|\Phi\rangle \|^2E(\alpha|\Psi\rangle+\beta|\Phi\rangle) 
\leq f(t)\;,
\end{equation}
for all $0\leq t\leq1$, where
$$
f(t)=\frac{t|\beta|^2+(1-t)|\alpha|^2}{t(1-t)}
\Big[tE(|\Psi\rangle)+(1-t)E(|\Phi\rangle)+h_2(t)\Big]\;.
$$
\end{theorem}
Comments: (\textbf{i}) For $t=|\alpha|^2$ we get the LPS bound; i.e.
$f(|\alpha|^2)=2[|\alpha|^2E(\Psi)+|\beta|^2E(\Phi)+h_2\left(|\alpha|^2\right)]$.\\
(\textbf{ii}) Note that $f(t)\geq |\alpha|^2E(\Psi)+|\beta|^2E(\Phi)+h_2\left(|\alpha|^2\right)$.
This is consistent with the case of biorthogonal states.\\
(\textbf{iii}) The minimum of the function $f(t)$ is obtained at $t=t^{*}$ where $t^{*}$
satisfies the implicit equation:
$$
\frac{|\alpha|^2(1-t^{*})^2}{|\beta|^2(t^{*})^2}
=\frac{E(\Psi)-\log t^{*}}{E(\Phi)-\log(1-t^{*})}\;.
$$
(\textbf{iv}) Using the same idea presented in theorem 2, the upper bound in theorem 3
can be improved a bit by replacing $h_2(t)$ in $f(t)$ with 
$h_2(t)-|S(\rho^{A}_{t})-S(\rho^{B}_{t})|$.\\ 
(\textbf{v}) For the trivial case where
$\alpha=1\;(\beta=0)$ we get $f(t)=E(\Psi)$ for $t=1$. 
That is, the upper bound equals $E(\Gamma)$. 
On the other hand, the LPS bound for $\alpha=1$ is $2E(\Psi)=2E(\Gamma)$.

\begin{proof}
Consider the state
$$
\rho^{AB}_{t}=t|\Psi\rangle\langle\Psi|+(1-t)|\Phi\rangle\langle\Phi|\;,
$$
where $0\leq t\leq 1$. For any $\theta, \phi\in[0,2\pi)$ we can construct a new 
decomposition of $\rho^{AB}_{t}$:
\begin{equation}
\rho^{AB}_{t}=q|\chi_{1}\rangle\langle\chi_{1}|+(1-q)|\chi_{2}\rangle\langle\chi_{2}| 
\label{newde}
\end{equation}
where 
\begin{align}
\sqrt{q}|\chi_{1}\rangle &
=\sqrt{t}\cos\theta|\Psi\rangle+\sqrt{1-t}e^{i\phi}\sin\theta|\Phi\rangle
\nonumber\\
\sqrt{1-q}|\chi_{2}\rangle &
=-\sqrt{t}e^{-i\phi}\sin\theta|\Psi\rangle+\sqrt{1-t}\cos\theta|\Phi\rangle
\nonumber\\
q &=\|\sqrt{t}\cos\theta|\Psi\rangle+\sqrt{1-t}e^{i\phi}\sin\theta|\Phi\rangle\|\;.
\label{q}
\end{align} 
Now, from the properties of the von-Neumann entropy given in 
Eqs.~(\ref{concave},\ref{conc}), we deduce that
\begin{align}
0 & \leq S(\rho^{A}_{t})-tE(|\Psi\rangle)-(1-t)E(|\Phi\rangle)\leq h_2(t)\nonumber\\
0 & \leq S(\rho^{A}_{t})-qE(|\chi_1\rangle)-(1-q)E(|\chi_2\rangle)\leq h_2(q)\;.\nonumber
\end{align}
From these inequalities we get
\begin{equation}
qE(|\chi_1\rangle)+(1-q)E(|\chi_2\rangle)\leq 
tE(|\Psi\rangle)+(1-t)E(|\Phi\rangle)+h_2(t)\;.
\end{equation}
Thus, since $E(|\chi_2\rangle)\geq 0$ we find that
\begin{equation}
E(|\chi_1\rangle)\leq\frac{1}{q}\Big[tE(|\Psi\rangle)+(1-t)E(|\Phi\rangle)+h_2(t)\Big]\;.
\label{bound}
\end{equation}
Note that so far we have 3 free parameters: $t$, $\theta$ and $\phi$.
We now concentrate on all the convex decompositions of $\rho^{AB}_{t}$
with $|\chi_1\rangle=|\Gamma\rangle/\|\Gamma\|$. This requirement reduces
the number of free parameters to one and can be expressed in terms of 
the following conditions (see Eq.~(\ref{q})):
\begin{equation}
\alpha'\equiv\frac{\alpha}{\|\Gamma\|}=\sqrt{\frac{t}{q}}\cos\theta\;\;\text{and}\;\;
\beta'\equiv\frac{\beta}{\|\Gamma\|}=\sqrt{\frac{1-t}{q}}e^{i\phi}\sin\theta 
\label{conditions}
\end{equation}
Since 
$|\Gamma\rangle=\alpha|\Psi\rangle+\beta|\Phi\rangle$ is defined up to 
a global phase we will assume, without loss of generality, that $\alpha$ is real
and non-negative. Similarly, we take $\phi$ to be equal to the phase of $\beta$
so that $|\beta'|=\sqrt{(t-1)/q}\sin\theta$. 
The parameter $q$ can be written as a function of $t$ and $\theta$. 
Note that 
$$
\frac{1}{\|\Gamma\|^2}=|\alpha'|^2+|\beta'|^2=\frac{t}{q}\cos^2\theta+\frac{1-t}{q}\sin^2\theta\;.
$$
Hence,
\begin{equation}
\frac{q}{\|\Gamma\|^2}=t\cos^2\theta+(1-t)\sin^2\theta\;.
\label{qqq}
\end{equation}
Now, substituting this form of $q$ into Eq.~(\ref{conditions}) provides the relation between
$t$ and $\theta$:
$$
\cos^2\theta=\frac{(1-t)|\alpha|^2}{t|\beta|^2+(1-t)|\alpha|^2}\;;\;
\sin^2\theta=\frac{t|\beta|^2}{t|\beta|^2+(1-t)|\alpha|^2}\;.
$$ 
Finally, using these relations in eq.~(\ref{qqq}) gives
$$
\frac{q}{\|\Gamma\|^2}=\frac{t(1-t)}{t|\beta|^2+(1-t)|\alpha|^2}\;.
$$
Hence, for decompositions with $|\chi_1\rangle=|\Gamma\rangle/\|\Gamma\|$
we get (see Eq.~(\ref{bound}))
\begin{align}
\|\Gamma\|^2 E(|\Gamma\rangle)& \leq
\frac{t|\beta|^2+(1-t)|\alpha|^2}{t(1-t)}\nonumber\\
&\times\Big[tE(|\Psi\rangle)+(1-t)E(|\Phi\rangle)+h_2(t)\Big]\;.\nonumber
\end{align}
for all $0< t < 1$.
\end{proof}

\begin{example}
Here we consider an example where both Alice and Bob have Hilbert spaces
of dimension $d$:
\begin{align}
&|\Psi\rangle = \frac{1}{\sqrt{2}}\left(|11\rangle
+\frac{1}{\sqrt{d-1}}\left[|22\rangle+|33\rangle+\cdots+|dd\rangle\right]\right)\;,\nonumber\\
&|\Phi\rangle = \frac{1}{\sqrt{2}}\left(|11\rangle
-\frac{1}{\sqrt{d-1}}\left[|22\rangle+|33\rangle+\cdots+|dd\rangle\right]\right)\;,\nonumber\\
&\alpha = \frac{3}{5}\;\;\text{and}\;\;\beta=-\frac{4}{5}\;.
\end{align}
This is the same example as one of the examples given in~\cite{Lin06} except
that here $\alpha\neq -\beta$. One can easily check that 
$E(\Psi)=E(\Phi)=\frac{1}{2}\log(d-1)+1$ and 
$E(\alpha|\Psi\rangle+\beta|\Phi\rangle)=(49/50)\log(d-1)+h_2(1/50)$.
Furthermore, it can be shown that in the limit $d\rightarrow\infty$ the minimum
of the function $f(t)$ is obtained at $t=3/7$. We therefore take this value
to get an upper bound $f(3/7)=(49/50)\log(d-1)+(49/25)h_2(3/7)$.
Thus, we have $f(3/7)-E(\Gamma)=O(1)$.
On the other hand, for large $d$ the LPS bound is approximately $\log(d-1)$
and so we have $\log(d-1)-E(\Gamma)\approx (1/50)\log(d-1)\rightarrow\infty$
as $d\rightarrow\infty$. That is, in the limit of high dimensions the LPS bound 
diverges from $E(\Gamma)$ 
whereas the our bound approaches $E(\Gamma)$.
\end{example}

We know move to discuss lower bounds.

\begin{theorem}
Let $|\Psi\rangle$ and $|\Phi\rangle$ be two bipartite states, and let
$|\Gamma\rangle=\alpha|\Psi\rangle+\beta|\Phi\rangle$ be a normalized
state (i.e. $\|\Gamma\|=1$) for some $\alpha,\beta\in\mathbb{C}$. Then,
\begin{equation}
E(|\Gamma\rangle)\geq\max\{L_1(t), L_2(t)\}\;, 
\end{equation}
for all $0\leq t\leq1$, where
\begin{align}
& L_1(t)=\frac{(1-t)|\beta|^2}{1-t(1-|\alpha|^2)}E(\Phi)
-\frac{1-t}{t}E(\Psi)-\frac{1}{t}h_2(t)\nonumber\\
& L_2(t)=\frac{(1-t)|\alpha|^2}{1-t(1-|\beta|^2)}E(\Psi)
-\frac{1-t}{t}E(\Phi)-\frac{1}{t}h_2(t)\nonumber.
\end{align}
\end{theorem}

Comments: (\textbf{i}) If $|\Psi\rangle$ and $|\Phi\rangle$ are orthogonal
(i.e. $|\alpha|^2+|\beta|^2=1$) then we can obtain a simple lower bound
by taking $t=1/(2|\alpha|^2)$ (or $t=1/(2|\beta|^2)$ if $|\alpha|^2<1/2$):
$$
E(|\Gamma\rangle)\geq (|\beta|^2-|\alpha|^2)\left[E(\Phi)-E(\Psi)\right]
-\frac{1}{|\gamma|^2}h_2(|\alpha|^2)\;,
$$
where $|\gamma|^2=\max\{|\alpha|^2,|\beta|^2\}$. Note that in general $t=1/(2|\alpha|^2)$
does not maximize the function $L_1(t)$ (or $L_1(t)$) and therefore the bound above is 
just a simple bound and \emph{not} the optimal one.\\
(\textbf{ii}) Note that it is inappropriate to replace $h_2(t)$ in the theorem above with 
$h_2(t)-|S(\rho^{A}_{t})-S(\rho^{B}_{t})|$ in order to improve it a bit. The reason
is that this time $\rho^{A}_{t}$ is a mixture that consists of $|\Gamma\rangle$
itself.\\
(\textbf{iii}) For $\alpha=0$ (or $\beta=0$) the lower bound is $E(\Gamma)$.
This gives us the first indication that the lower bound is tight.\\
(\textbf{iv}) The maximum of the function $L_1(t)$ is obtained at $t=t^{*}$ where $t^{*}$
satisfies the implicit equation:
\begin{equation}
\frac{|\alpha|^2|\beta|^2(t^{*})^2}{\left[1-(1-|\alpha|^2)t^{*}\right]^2}E(\Phi)
=E(\Psi)-\log(1-t^{*})\;.\label{max}
\end{equation}
Similar expression can be found for the value of $t$ that maximizes $L_2(t)$.

\begin{proof}
Consider the state
$$ 
\rho^{AB}_{t}=t|\Gamma\rangle\langle\Gamma|+(1-t)|\Psi\rangle\langle\Psi|\;.
$$
where $0\leq t\leq 1$. For any $\theta, \phi\in[0,2\pi)$ we can construct a new 
decomposition of $\rho^{AB}_{t}$ just as in $Eq.(\ref{newde})$ except that now:
\begin{align}
\sqrt{q}|\chi_{1}\rangle &
=\sqrt{t}\cos\theta|\Gamma\rangle+\sqrt{1-t}e^{i\phi}\sin\theta|\Psi\rangle
\nonumber\\
\sqrt{1-q}|\chi_{2}\rangle &
=-\sqrt{t}e^{-i\phi}\sin\theta|\Gamma\rangle+\sqrt{1-t}\cos\theta|\Psi\rangle
\nonumber\\
q &=\|\sqrt{t}\cos\theta|\Gamma\rangle+\sqrt{1-t}e^{i\phi}\sin\theta|\Psi\rangle\|\;.
\label{www}
\end{align} 
Using the same arguments as in Theorem 3 we find that:
$$
qE(|\chi_1\rangle)+(1-q)E(|\chi_2\rangle)\leq 
tE(|\Gamma\rangle)+(1-t)E(|\Psi\rangle)+h_2(t)\;.
$$
Now, since $E(|\chi_2\rangle)\geq 0$ we have
\begin{equation}
E(|\Gamma\rangle)\geq\frac{1}{t}\left[qE(|\chi_1\rangle)-(1-t)E(|\Psi\rangle)-h_2(t)\right]\;.
\label{lb}
\end{equation}
This equation holds for any choice of $\theta$ and $\phi$. We would like now
to choose $\theta$ and $\phi$ such that $|\chi_1\rangle=|\Phi\rangle$. From
Eq.(\ref{www}) we find that it is possible if 
$$
\sqrt{q}=\beta\sqrt{t}\cos\theta\;\;\text{and}\;\;\sqrt{t}\alpha\cos\theta
=-\sqrt{1-t}e^{i\phi}\sin\theta\;.
$$
Without loss of generality, we can assume that $\beta$ is real (since $|\Gamma\rangle$
is defined up to a global phase) and we take $-e^{i\phi}$ to be the phase of $\alpha$.
Furthermore, from these equations it follows that
$$
q=\frac{t(1-t)|\beta|^2}{1-t(1-|\alpha|^2)}\;.
$$
Substituting this value for $q$ in Eq.~(\ref{lb}) gives the lower bound $L_1(t)$.
The lower bound $L_2(t)$ is similarly obtained by exchanging the roles of $|\Psi\rangle$
and $|\Phi\rangle$.
\end{proof}

In the following example we show that our lower bound can be very tight.
\begin{example}
Here we take $|\Psi\rangle$, $|\Phi\rangle$ and $\alpha$
to be exactly the same as in Example 3, and $\beta=4/5$. We therefore have
$E(\Psi)=E(\Phi)=\frac{1}{2}\log(d-1)+1$ whereas
$E(\alpha|\Psi\rangle+\beta|\Phi\rangle)=(1/50)\log(d-1)+h_2(1/50)$.
We would like now to find the value $t=t^{*}$ in Eq.~(\ref{max}) at which $L_1(t)$
is maximum. In the limit $d\rightarrow\infty$ we can ignore the logarithmic term
in Eq.~(\ref{max}) and so we get $t^{*}=25/28$. The value of $L_1(t)$ at $t=25/28$
is $L_{1}(25/28)=(1/50)\log(d-1)+(1/25)-(28/25)h_2(25/28)$. We therefore get 
that $L_1(25/28)/E(\alpha|\Psi\rangle+\beta|\Phi\rangle)\rightarrow 1$ at the limit $d\rightarrow\infty$ and
$E(\alpha|\Psi\rangle+\beta|\Phi\rangle)-L_1(25/28)\approx 0.65$ . The last value
can be improved if one takes into account the logarithmic term in Eq.~(\ref{max}).
\end{example}

We end by making two observations.
First, the lower bound given in Theorem 4 can also provide
a lower bound on the entanglement of 2-dimensional bipartite
subspaces~\cite{Gou07} (see also~\cite{Cub07} for the Schmidt rank
of subspaces) by minimizing the bound over $\alpha$ 
and $\beta$. This minimization also provides a lower bound on 
the entanglement of formation of a density matrix whose support subspace
is spanned by $|\Psi\rangle$ and $|\Phi\rangle$. 
Second, in this Letter we have given lower and upper bounds
for the entanglement of superpositions including two states. The question
regarding the entanglement of superpositions with more than two terms
is an important one for future work. 

\emph{Acknowledgments:---}
The author acknowledges financial support from NSERC.

\end{document}